\documentclass[%
 reprint,
superscriptaddress,
 amsmath,amssymb,
 aps,
floatfix,
]{revtex4-2}

\usepackage{graphicx}
\usepackage{dcolumn}
\usepackage{bm}

\begin{document}

\title{Optically induced static magnetic field in ensemble of \\nitrogen-vacancy centers in diamond}

\author{Farid Kalhor}
\author{Noah F. Opondo}
\author{Shoaib Mahmud}
\author{Leif Bauer}
\affiliation{The Elmore Family School of Electrical and Computer Engineering, Purdue University, West Lafayette, IN 47906, USA}%

\author{Li-Ping Yang}
\affiliation{Center for Quantum Sciences and School of Physics, Northeast Normal University, Changchun 130024, China}%

\author{Sunil A. Bhave}
\author{Zubin Jacob}
 \email{zjacob@purdue.edu}
\affiliation{The Elmore Family School of Electrical and Computer Engineering, Purdue University, West Lafayette, IN 47906, USA}%

\begin{abstract}
Generation of local magnetic field at the nanoscale is desired for many applications such as spin-qubit-based quantum memories. However, this is a challenge due to the slow decay of static magnetic fields. Here, we demonstrate photonic spin density (PSD) induced effective static magnetic field for an ensemble of nitrogen-vacancy (NV) centers in bulk diamond.  This locally induced magnetic field is a result of coherent interaction between the optical excitation and the NV centers. We demonstrate an optically induced spin rotation on the Bloch sphere exceeding 10 degrees which has potential applications in all optical coherent control of spin qubits.
\end{abstract}

\maketitle

Optically inducing a magnetic field is a promising route for generating high gradient magnetic fields and nanoscale on-chip excitations with a fast temporal response. For example, magneto-optical Barnett effect uses circularly polarized light to switch magnetization in magnetic thin films using ultrafast optical pulses~\cite{rebei_magneto-optical_2008,nakata_optomagnonic_2020}, valley-selective optical pumping can induce nonreciprocity in transition metal dichalcogenides~\cite{guddala_all-optical_2021}, while structured light can be used to induce a magnetic field by shaping optically induced currents ~\cite{jana_reconfigurable_2021}. Recently it was shown that a single nitrogen-vacancy (NV) center can probe an effective static magnetic field induced by the photonic spin density (PSD) of an optical beam~\cite{kalhor_quantum_2021}. PSD is the spatial distribution of the degree of circular polarization of an optical field at the nanoscale.
While most optically induced magnetic fields rely on absorption of light and are incoherent interactions, the optically induced magnetic field probed by a single NV center is non-absorbing and coherently interacts with the NV center even at room temperature~\cite{kalhor_quantum_2021}.

\begin{figure*}[tbh!]
\centering
\includegraphics[width=160mm]{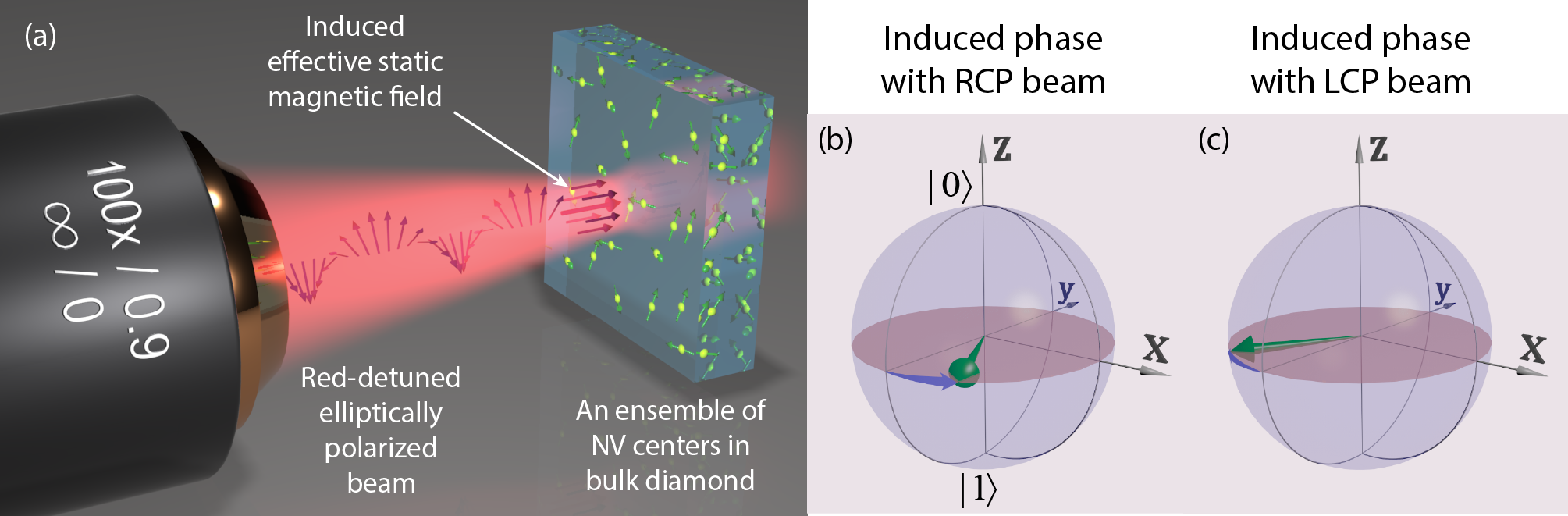}
\caption{Interaction of the PSD of an optical beam with the electronic spin of NV centers. (a) Schematic of the measurements where a focused beam with elliptical polarization interacts with an ensemble of NV centers. (b) The Bloch sphere representation of the NV centers' electronic spin showing an induced rotation caused by the optical beam. The amount of this induced rotation depends on the PSD of the beam. The direction of the PSD and the bias magnetic field are aligned such that the induced rotations are equal for all NV centers in the ensemble.}
\label{fig1}
\end{figure*}

The NV center in diamond has emerged as a single-spin quantum magnetometer for probing condensed matter phenomena \cite{ku_imaging_2020, thiel_probing_2019, abobeih_atomic-scale_2019, mclaughlin_strong_2021}. The recent work by Kalhor et al. demonstrating the optically induced static magnetic field for the NV center shows that this quantum sensor can also be used to probe properties of light fields at the nanoscale~\cite{kalhor_quantum_2021}. 
This work uses a single NV center on an atomic force microscopy (AFM) tip to measure the optically induced effective static magnetic field. 
While reference~\cite{kalhor_quantum_2021} focuses on probing the photonic spin density (PSD), the demonstrated induced magnetic field can also be used for coherent control of NV center qubits. This coherent control of spin qubits is arguably a consequential application for the PSD based phenomenon.
A more suitable platform to study the coherent control of NV center qubit is NV centers in bulk diamond which can be used for on-chip quantum technologies.
Moreover, in a bulk diamond with planar interface the optical fields experience a minimal distortion as opposed to the AFM tip. The nanoscale structure of the tip distorts the optical fields and makes a quantitative study of this PSD based effect and its dependence on the wavelength of the excitation beam very challenging. It is also crucial to know whether the optically induced static magnetic field can be observed with an ensemble of NV centers or the random orientation of the NV centers in the ensemble leads to a zero net effect. 

In this work we demonstrate room-temperature optically induced static magnetic field for ensemble of NV centers in bulk diamond. We show that the interaction of this effective magnetic field with NV centers leads to rotation of the spin qubits on the Bloch sphere proportional to the PSD of the optical beam and the interaction time between the optical beam and the spin qubit. To maximize the interaction time, we use the Hahn echo technique to increase the coherence time of the spin qubits. We show that the strength of the field is inversely dependent on the detuning between the frequency of the optical beam and the optical transition of NV centers. We further study the probability of off-resonant absorption of the optical beam by the NV centers to ensure low impact on the coherence of the spin qubits. We show that the optically induced static magnetic field is a broadband effect. Finally, we study the wavelength dependence of the effective static magnetic field to explore the optimal regimes of optical qubit control. The results of this work will help pave the way for enhancements in the effective static magnetic field and expanding its range of applications to on-chip spin quantum electrodynamics and all-optical coherent spin qubit control.

Fig.~\ref{fig1} shows the schematics of the experiment. An elliptically polarized beam is focused onto a diamond sample with an ensemble of NV centers. The beam possesses PSD, related to the ellipticity of its polarization \cite{berry_optical_2009, kalhor_universal_2016, yang_quantum_2020}. The electric PSD, in the monochromatic limit, is defined as 
$\vec{S}^{\rm obs}_E=(\epsilon/4\omega_0){\rm Im}(\vec{E}^*\times\vec{E})$ where $\epsilon$ is the permittivity of the medium and $\vec{E}^{*}$ denotes the complex conjugate of the complex electric field.
The interaction of the beam with NV centers induces a rotation in the spin vector of the NV centers on the Bloch sphere. This rotation is proportional to the PSD of the beam and we define an effective static magnetic field to describe the strength of this interaction (see \cite{kalhor_quantum_2021}). The NV centers are randomly scattered in the bulk of a single crystalline diamond, with their four different orientations determined by the (100) growth plane. A bias magnetic field is applied normal to the surface of the sample and is aligned to have the same projection along all NV center orientations. The PSD of the elliptically polarized beam is parallel to the direction of propagation and is also normal to the surface of the sample. This use of symmetries in the system ensures that the PSD induced rotation is the same over all of the NV centers. Moreover, the spot size of the PSD beam is designed to be larger than the spot size of the excitation laser to ensure that all excited NV centers are illuminated uniformly with the PSD beam. Figures \ref{fig1}(b) and \ref{fig1}(c) illustrate the rotation of the spin vector on the Bloch sphere. These two figures show the rotations induced by right-handed (RCP) and left-handed (LCP) circularly polarized incident beams. The two different polarizations induce rotations in opposite directions due to their different PSD. We measure these induced rotations by performing ac magnetometry using the Hahn echo technique.

\begin{figure*}[tbh!]
\centering
\includegraphics[width=160mm]{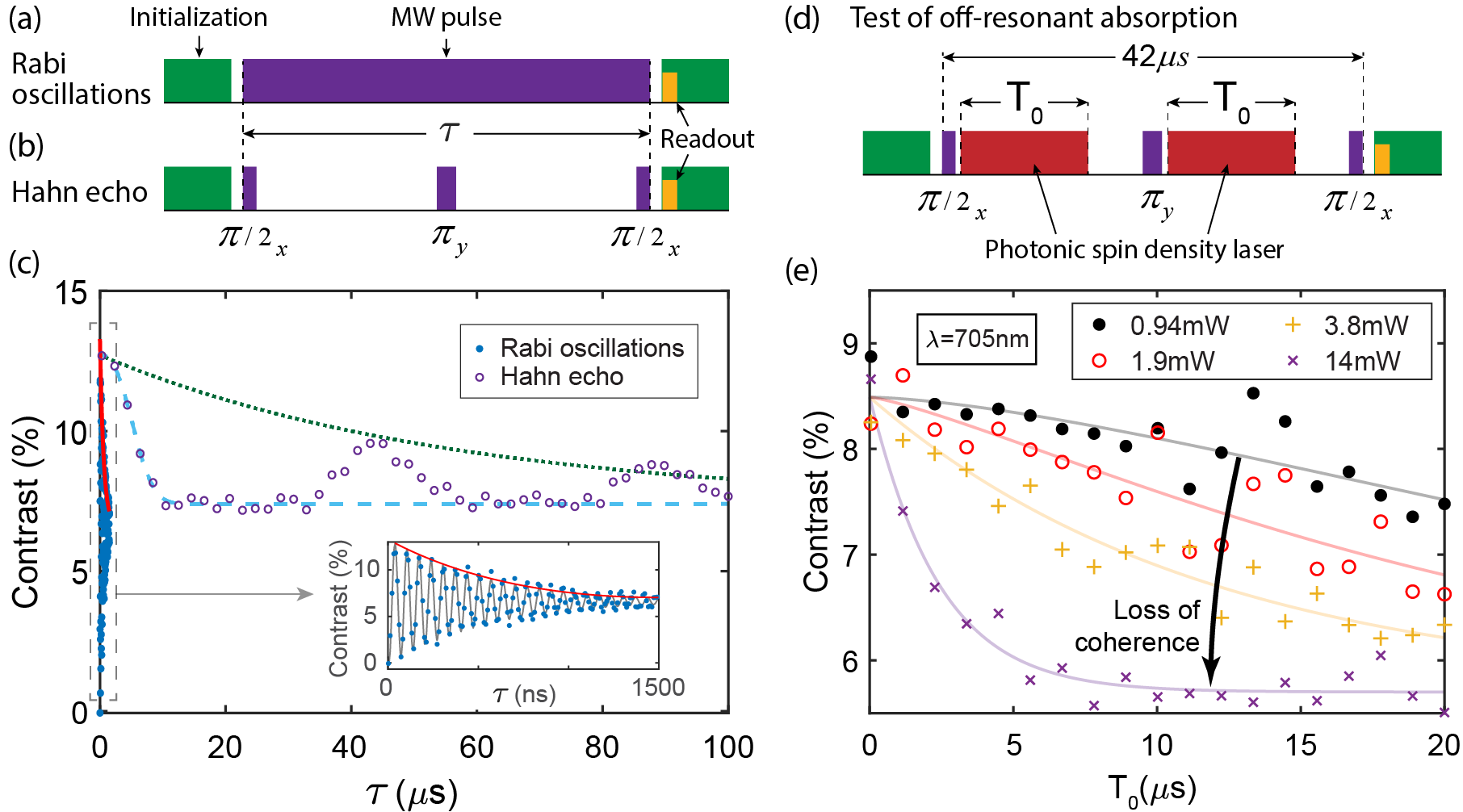}
\caption{Optimizing the coherence time of NV center ensemble.
(a) The pulse sequence used for measuring Rabi oscillations and inhomogeneous broadening $T_2^*$ for the ensemble.
(b) The Hahn echo pulse sequence for measuring spin echo and coherence time $T_2$.
(c) Measurement results for the pulse sequences shown in (a) (blue dots) and (b) (purple circles). The Rabi oscillations show the shortest timescale, $T_2^*$. The inset shows the details of this measurement. The red line is a fit to the envelope of the Rabi oscillations. The Hahn echo signal shows a slower decay due to elimination of inhomogeneous broadening and periodic revivals in the signal. The fits to the decay and revivals are shown with dashed blue and dotted green lines, respectively. The revivals are due to the periodic motion of the carbon-13 nuclear spins under the bias magnetic field. We use the second revival point ($\tau=42\mu s$) for the effective static magnetic field measurements.
(d) A pulse sequence designed to measure the decoherence caused by the off-resonant absorption of the PSD signal. Here we add the PSD signal to both halves of the Hahn echo sequence to avoid spin rotations due to the effective static magnetic field. By increasing the pulse length $T_0$ we increase the probability of off-resonant absorption events in one measurement and observe the trend of the loss of the coherence signal.
(e) The decoherence induced by the PSD signal for different PSD laser powers measured for $\lambda=705nm$. As the power of the PSD beam increases, the off-resonant absorption increases and results in decoherence of the spin-qubit.
}
\label{fig2}
\end{figure*}

\begin{figure*}[tbh!]
\centering\includegraphics[width=140mm]{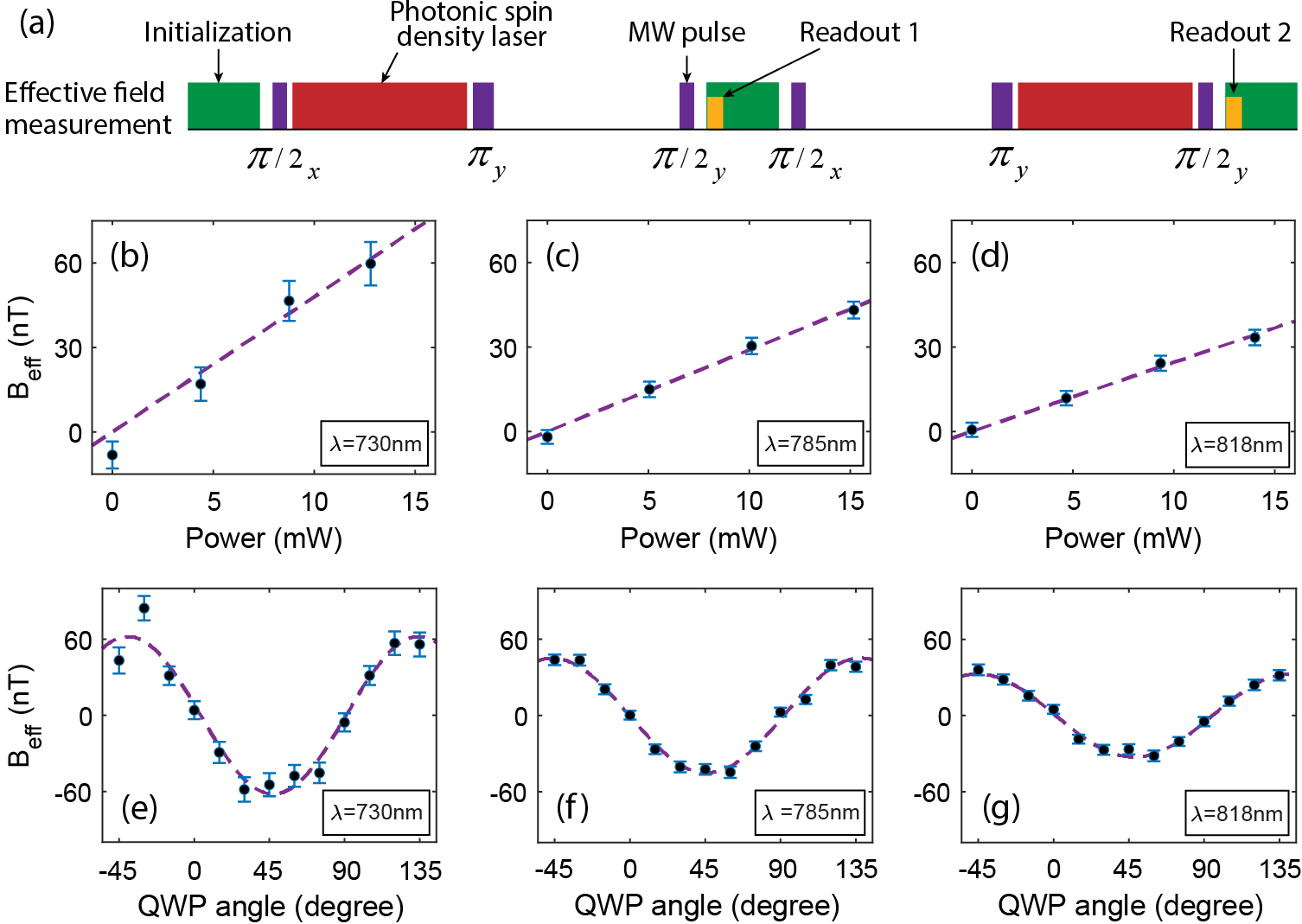}
\caption{Scaling  of the effective static magnetic field with beam power and polarization ellipticity.
(a) The PSD signal is added to the Hahn echo pulse sequence. Since this pulse sequence is only sensitive to ac magnetic fields, we add the PSD signal to one half of the sequence only. Adding the PSD pulse to the first or second half induces rotations in opposite directions in the spin vector of the NV center. Therefore, we perform two measurements-- with PSD signal added to the first or second half of the Hahn echo sequence-- and calculate the amplitude of the effective static magnetic field by subtracting the outcome of these two measurements.
(b)-(d) Scaling of the effective field with the optical power of the PSD beam for different wavelengths. We observe linear dependence, independent of the wavelength.
(e)-(g) The dependence of the effective field on the ellipticity of polarization for different wavelengths. The ellipticity is controlled by rotating a QWP by an angle $\theta$ where $\theta=0^\circ , 90^\circ$ correspond to linear polarization and $\theta=\pm45^\circ , 135^\circ$ correspond to circular polarization. The fits to the curves are $y=sin(2\theta)$ which is the expected dependence of PSD to QWP angle $\theta$. We observe good agreement between theory and experiment for all wavelengths. Error bars show one standard deviation.
}
\label{fig4}
\end{figure*}

Figure~\ref{fig2} shows the pulse sequence used for the ac magnetometry measurements and our approach to optimizing the sensitivity of the measurement. The sensitivity is proportional to $\eta \propto \frac{1}{\sqrt{T_2}}$ \cite{taylor_high-sensitivity_2008}.
Here $\eta$ has units of $nT / \sqrt{Hz}$ and $T_2$ is the NV centers' coherence time. Therefore, to optimize the sensitivity we maximize the coherence time. Figure~\ref{fig2}(a) shows the pulse sequence used for measuring Rabi oscillations. The outcome of this measurement is shown in Fig.~\ref{fig2}(c) and its inset. The red curve in this figure shows a fit to the envelope of the oscillations. The decay of this curve determines the coherence time $T_2^*$ of the spin qubits. This coherence time is limited by inhomogeneous broadening of the ensemble resonance. This broadening is a result of the background magnetic noise generated by carbon-13 atoms in the diamond crystal. To reduce the effect of this noise we use Hahn echo pulse sequence shown in Fig.~\ref{fig2}(b) \cite{taylor_high-sensitivity_2008, naydenov_dynamical_2011}. This pulse sequence takes advantage of the slowly varying nature of the background noise to lower its effect on the spin qubits. The outcome of this measurement is shown in Fig.~\ref{fig2}(c). The decay of this signal (blue dashed line) is slower than the decay of the Rabi oscillations due to the lower effective noise. Moreover, there are revivals in the signal as a result of the periodic motion of nuclear spins under the influence of the applied bias magnetic field \cite{childress_coherent_2006}. The envelope of the periodic revivals (dotted green line) shows the slowest decay in the coherence signal. We use the second revival ($\tau \approx 42 \mu s$) for our measurements to achieve a long interaction time and the highest sensitivity. 

Illuminating the NV centers with the PSD laser can affect their coherence through off-resonant absorption. To capture this effect we use the pulse sequence shown in Fig.~\ref{fig2}(d). In this pulse sequence the rotations induced by the two PSD pulses cancel each other and we only see decoherence caused by the addition of the PSD laser. By increasing the pulse length $\tau$ we can study this effect. Figure~\ref{fig2}(e) shows the decoherence effect from off-resonant absorption for different laser powers. The four sets of data show how different laser powers affect the measurement outcome for $\lambda=705nm$. The x-axis shows the length of each of the PSD pulses. As the pulse length increases, the probability of off-resonant absorption also increases. The results of this measurement helps us determine the maximum laser power which does not cause significant decoherence to the NV centers. It should be mentioned that we do not observe any decoherence for wavelengths $\lambda=785nm$ and $\lambda=818nm$ with laser powers reaching $P=20mW$. As the wavelength of the PSD laser gets closer to the optical transition of the NV center at $\lambda=637nm$, off-resonant absorption becomes stronger. The decoherence caused by this effect is noticeable for $\lambda=730nm$ and becomes significant for $\lambda=705nm$ as shown in Fig.~\ref{fig2}(e). 

The pulse sequence for measuring the effective static magnetic field is shown in Fig.~\ref{fig4}(a). The sequences consists of two measurements to take advantage of the phase of an ac signal: if the PSD pulse is sent in the first half of the Hahn echo sequence (first measurement) it induces a positive phase in the spin qubit; if it is sent in the second half (second measurement) the induced phase will have the opposite sign. Therefore, we perform two measurements as shown in Fig.~\ref{fig4}(a) and subtract the outcome of the two measurements to calculate the effective static magnetic field induced by the PSD signal. This technique helps with eliminating possible systematic errors in the readout process. 

Figures~\ref{fig4}(b)-(g) depict the outcome of the effective static magnetic field measurements. The effective static magnetic field is proportional to the PSD of the optical beam. We can tune the PSD of the optical beam by controlling its power and its polarization. The PSD of the optical beam is linearly proportional to the optical power. Figures~\ref{fig4}(b)-(d) show how the effective field depends on the optical power of the PSD beam for three different wavelengths. The dashed lines show linear fits to the data. The linear dependence of the effective field on the optical power matches the expected behavior for PSD.

Figures~\ref{fig4}(e)-(g) show the dependence of the effective static magnetic field on the polarization of the PSD beam for three different wavelengths. The beam polarization is controlled with a quarter-wave plate (QWP). The PSD of the beam has a $\vec{S}^{\rm obs}_E \propto sin(2\theta) \hat{z}$ dependence on the rotation angle of the QWP ($\theta$) where $\hat{z}$ is the direction of propagation of the beam. The dashed lines show a fit of $sin(2\theta)$ to the data for each wavelength. Figure~\ref{fig4} shows that the measured effective static magnetic field is indeed proportional to the PSD for different wavelengths. The optically induced effective magnetic field reaches $B_{eff}=60nT$ which corresponds to a spin rotation of $\phi\approx12^{\circ}$. 

\begin{figure*}[tbh!]
\centering\includegraphics[width=75mm]{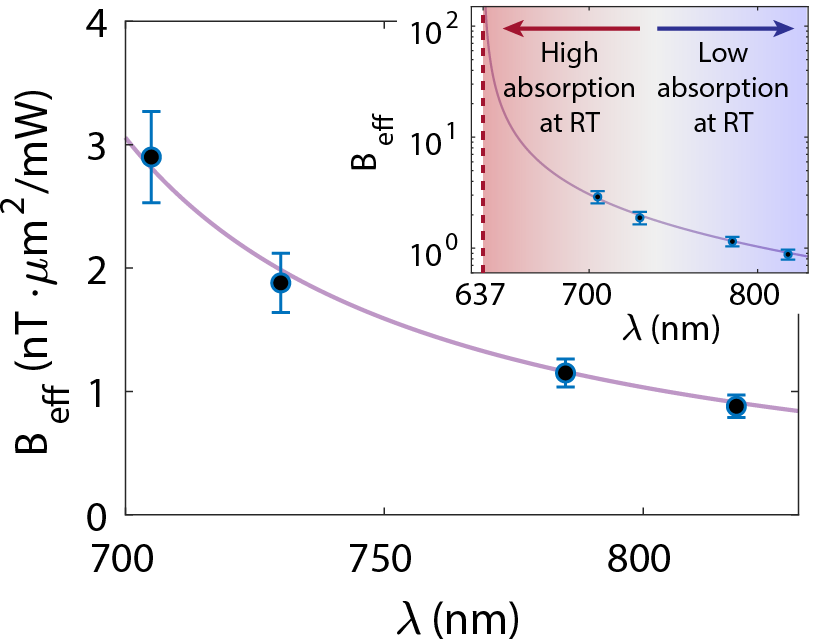}
\caption{The dependence of the effective static magnetic field strength on the wavelength of the PSD beam. The rapid increase of field strength as the wavelength decreases shows a promising route for enhancing the effective static magnetic field. The inset shows the data in log scale and the optical transition of the NV center at $\lambda=637nm$. As the detuning decreases, the off-resonant absorption by the NV centers affects their coherence and the error bars become larger. For each data point two sets of measurements are performed with QWP angles $\theta=\pm45^\circ$. The data points show the average between the amplitude of the two measurements normalized to the power density of the beams. Error bars show one standard deviation.}
\label{fig5}
\end{figure*}

The main goal of this work is to determine the wavelength dependence of the effective static magnetic field in the far off-resonant limit. This result is shown in Fig.~\ref{fig5} where the measurements are performed for four different wavelengths. The main factor that affects the strength of the effective static magnetic field is the detuning between the laser wavelength and the optical transition of the NV centers. As the detuning decreases, the probability of off-resonant absorption by the NV centers increases. This absorption leads to decoherence of the NV centers and loss of sensitivity in the measurement (Fig.~\ref{fig2}(e)). In these measurements the off-resonant absorption is noticeable for $\lambda=730nm$ and becomes a dominant factor at $\lambda=705nm$. In order to measure the effective static magnetic field for $\lambda=705nm$ we have cooled down the sample to $T=265K$ using a thermoelectric cooler. 

The purple line in Fig.~\ref{fig5} shows a fit to the data $y=C/\lambda\Delta$ where $C$ is a constant and $\Delta$ is  the detuning. The inset shows the data on log-scale with the optical transition of the NV centers at $\lambda=637nm$. The fast growing nature of the effective static magnetic field for smaller detunings is captured in the inset. However, due to the off-resonant absorption, it is not possible to probe the system for wavelengths smaller than $\lambda=700nm$ at room temperature. Unraveling the scaling of the effective static magnetic field for near-resonant wavelengths requires rigorous future studies at cryogenic temperatures.

In conclusion, we have demonstrated the optically induced static magnetic field for an ensemble of NV centers at room temperature. We show that this effective static magnetic field is proportional to the PSD of the off-resonant excitation. The decoherence induced by the off-resonant optical beam is studied for different wavelengths. Using this platform, we have characterized the wavelength dependence of the optically induced static magnetic field in the far-off-resonant regime. The results of this study can pave the way to utilize this effective magnetic field for on-chip applications and all-optical control of qubits.

\begin{acknowledgments}
Authors acknowledge funding from DARPA Nascent Light-Matter Interactions (NLM) and Army Research Office (W911NF-21-1-0287). Leif Bauer acknowledges the National Science Foundation for support under the Graduate Research Fellowship Program (GRFP) under grant number DGE-1842166.
\end{acknowledgments}

\bibliography{1-Refs}

\begin{thebibliography}{15}%
\makeatletter
\providecommand \@ifxundefined [1]{%
 \@ifx{#1\undefined}
}%
\providecommand \@ifnum [1]{%
 \ifnum #1\expandafter \@firstoftwo
 \else \expandafter \@secondoftwo
 \fi
}%
\providecommand \@ifx [1]{%
 \ifx #1\expandafter \@firstoftwo
 \else \expandafter \@secondoftwo
 \fi
}%
\providecommand \natexlab [1]{#1}%
\providecommand \enquote  [1]{``#1''}%
\providecommand \bibnamefont  [1]{#1}%
\providecommand \bibfnamefont [1]{#1}%
\providecommand \citenamefont [1]{#1}%
\providecommand \href@noop [0]{\@secondoftwo}%
\providecommand \href [0]{\begingroup \@sanitize@url \@href}%
\providecommand \@href[1]{\@@startlink{#1}\@@href}%
\providecommand \@@href[1]{\endgroup#1\@@endlink}%
\providecommand \@sanitize@url [0]{\catcode `\\12\catcode `\$12\catcode
  `\&12\catcode `\#12\catcode `\^12\catcode `\_12\catcode `\%12\relax}%
\providecommand \@@startlink[1]{}%
\providecommand \@@endlink[0]{}%
\providecommand \url  [0]{\begingroup\@sanitize@url \@url }%
\providecommand \@url [1]{\endgroup\@href {#1}{\urlprefix }}%
\providecommand \urlprefix  [0]{URL }%
\providecommand \Eprint [0]{\href }%
\providecommand \doibase [0]{https://doi.org/}%
\providecommand \selectlanguage [0]{\@gobble}%
\providecommand \bibinfo  [0]{\@secondoftwo}%
\providecommand \bibfield  [0]{\@secondoftwo}%
\providecommand \translation [1]{[#1]}%
\providecommand \BibitemOpen [0]{}%
\providecommand \bibitemStop [0]{}%
\providecommand \bibitemNoStop [0]{.\EOS\space}%
\providecommand \EOS [0]{\spacefactor3000\relax}%
\providecommand \BibitemShut  [1]{\csname bibitem#1\endcsname}%
\let\auto@bib@innerbib\@empty
\bibitem [{\citenamefont {Rebei}\ and\ \citenamefont
  {Hohlfeld}(2008)}]{rebei_magneto-optical_2008}%
  \BibitemOpen
  \bibfield  {author} {\bibinfo {author} {\bibfnamefont {A.}~\bibnamefont
  {Rebei}}\ and\ \bibinfo {author} {\bibfnamefont {J.}~\bibnamefont
  {Hohlfeld}},\ }\bibfield  {title} {\bibinfo {title} {The magneto-optical
  {Barnett} effect: {Circularly} polarized light induced femtosecond
  magnetization reversal},\ }\href@noop {} {\bibfield  {journal} {\bibinfo
  {journal} {Physics Letters A}\ }\textbf {\bibinfo {volume} {372}},\ \bibinfo
  {pages} {1915} (\bibinfo {year} {2008})}\BibitemShut {NoStop}%
\bibitem [{\citenamefont {Nakata}\ and\ \citenamefont
  {Takayoshi}(2020)}]{nakata_optomagnonic_2020}%
  \BibitemOpen
  \bibfield  {author} {\bibinfo {author} {\bibfnamefont {K.}~\bibnamefont
  {Nakata}}\ and\ \bibinfo {author} {\bibfnamefont {S.}~\bibnamefont
  {Takayoshi}},\ }\bibfield  {title} {\bibinfo {title} {Optomagnonic {Barnett}
  effect},\ }\href@noop {} {\bibfield  {journal} {\bibinfo  {journal} {Physical
  Review B}\ }\textbf {\bibinfo {volume} {102}},\ \bibinfo {pages} {094417}
  (\bibinfo {year} {2020})}\BibitemShut {NoStop}%
\bibitem [{\citenamefont {Guddala}\ \emph {et~al.}(2021)\citenamefont
  {Guddala}, \citenamefont {Kawaguchi}, \citenamefont {Komissarenko},
  \citenamefont {Kiriushechkina}, \citenamefont {Vakulenko}, \citenamefont
  {Chen}, \citenamefont {Alù}, \citenamefont {M.~Menon},\ and\ \citenamefont
  {Khanikaev}}]{guddala_all-optical_2021}%
  \BibitemOpen
  \bibfield  {author} {\bibinfo {author} {\bibfnamefont {S.}~\bibnamefont
  {Guddala}}, \bibinfo {author} {\bibfnamefont {Y.}~\bibnamefont {Kawaguchi}},
  \bibinfo {author} {\bibfnamefont {F.}~\bibnamefont {Komissarenko}}, \bibinfo
  {author} {\bibfnamefont {S.}~\bibnamefont {Kiriushechkina}}, \bibinfo
  {author} {\bibfnamefont {A.}~\bibnamefont {Vakulenko}}, \bibinfo {author}
  {\bibfnamefont {K.}~\bibnamefont {Chen}}, \bibinfo {author} {\bibfnamefont
  {A.}~\bibnamefont {Alù}}, \bibinfo {author} {\bibfnamefont {V.}~\bibnamefont
  {M.~Menon}},\ and\ \bibinfo {author} {\bibfnamefont {A.~B.}\ \bibnamefont
  {Khanikaev}},\ }\bibfield  {title} {\bibinfo {title} {All-optical
  nonreciprocity due to valley polarization pumping in transition metal
  dichalcogenides},\ }\href@noop {} {\bibfield  {journal} {\bibinfo  {journal}
  {Nature Communications}\ }\textbf {\bibinfo {volume} {12}},\ \bibinfo {pages}
  {3746} (\bibinfo {year} {2021})}\BibitemShut {NoStop}%
\bibitem [{\citenamefont {Jana}\ \emph {et~al.}(2021)\citenamefont {Jana},
  \citenamefont {Herperger}, \citenamefont {Kong}, \citenamefont {Mi},
  \citenamefont {Zhang}, \citenamefont {Corkum},\ and\ \citenamefont
  {Sederberg}}]{jana_reconfigurable_2021}%
  \BibitemOpen
  \bibfield  {author} {\bibinfo {author} {\bibfnamefont {K.}~\bibnamefont
  {Jana}}, \bibinfo {author} {\bibfnamefont {K.~R.}\ \bibnamefont {Herperger}},
  \bibinfo {author} {\bibfnamefont {F.}~\bibnamefont {Kong}}, \bibinfo {author}
  {\bibfnamefont {Y.}~\bibnamefont {Mi}}, \bibinfo {author} {\bibfnamefont
  {C.}~\bibnamefont {Zhang}}, \bibinfo {author} {\bibfnamefont {P.~B.}\
  \bibnamefont {Corkum}},\ and\ \bibinfo {author} {\bibfnamefont
  {S.}~\bibnamefont {Sederberg}},\ }\bibfield  {title} {\bibinfo {title}
  {Reconfigurable electronic circuits for magnetic fields controlled by
  structured light},\ }\href@noop {} {\bibfield  {journal} {\bibinfo  {journal}
  {Nature Photonics}\ }\textbf {\bibinfo {volume} {15}},\ \bibinfo {pages}
  {622} (\bibinfo {year} {2021})}\BibitemShut {NoStop}%
\bibitem [{\citenamefont {Kalhor}\ \emph {et~al.}(2021)\citenamefont {Kalhor},
  \citenamefont {Yang}, \citenamefont {Bauer},\ and\ \citenamefont
  {Jacob}}]{kalhor_quantum_2021}%
  \BibitemOpen
  \bibfield  {author} {\bibinfo {author} {\bibfnamefont {F.}~\bibnamefont
  {Kalhor}}, \bibinfo {author} {\bibfnamefont {L.-P.}\ \bibnamefont {Yang}},
  \bibinfo {author} {\bibfnamefont {L.}~\bibnamefont {Bauer}},\ and\ \bibinfo
  {author} {\bibfnamefont {Z.}~\bibnamefont {Jacob}},\ }\bibfield  {title}
  {\bibinfo {title} {Quantum sensing of photonic spin density using a single
  spin qubit},\ }\href {https://doi.org/10.1103/PhysRevResearch.3.043007}
  {\bibfield  {journal} {\bibinfo  {journal} {Physical Review Research}\
  }\textbf {\bibinfo {volume} {3}},\ \bibinfo {pages} {043007} (\bibinfo {year}
  {2021})}\BibitemShut {NoStop}%
\bibitem [{\citenamefont {Ku}\ \emph {et~al.}(2020)\citenamefont {Ku},
  \citenamefont {Zhou}, \citenamefont {Li}, \citenamefont {Shin}, \citenamefont
  {Shi}, \citenamefont {Burch}, \citenamefont {Anderson}, \citenamefont
  {Pierce}, \citenamefont {Xie}, \citenamefont {Hamo}, \citenamefont {Vool},
  \citenamefont {Zhang}, \citenamefont {Casola}, \citenamefont {Taniguchi},
  \citenamefont {Watanabe}, \citenamefont {Fogler}, \citenamefont {Kim},
  \citenamefont {Yacoby},\ and\ \citenamefont {Walsworth}}]{ku_imaging_2020}%
  \BibitemOpen
  \bibfield  {author} {\bibinfo {author} {\bibfnamefont {M.~J.~H.}\
  \bibnamefont {Ku}}, \bibinfo {author} {\bibfnamefont {T.~X.}\ \bibnamefont
  {Zhou}}, \bibinfo {author} {\bibfnamefont {Q.}~\bibnamefont {Li}}, \bibinfo
  {author} {\bibfnamefont {Y.~J.}\ \bibnamefont {Shin}}, \bibinfo {author}
  {\bibfnamefont {J.~K.}\ \bibnamefont {Shi}}, \bibinfo {author} {\bibfnamefont
  {C.}~\bibnamefont {Burch}}, \bibinfo {author} {\bibfnamefont {L.~E.}\
  \bibnamefont {Anderson}}, \bibinfo {author} {\bibfnamefont {A.~T.}\
  \bibnamefont {Pierce}}, \bibinfo {author} {\bibfnamefont {Y.}~\bibnamefont
  {Xie}}, \bibinfo {author} {\bibfnamefont {A.}~\bibnamefont {Hamo}}, \bibinfo
  {author} {\bibfnamefont {U.}~\bibnamefont {Vool}}, \bibinfo {author}
  {\bibfnamefont {H.}~\bibnamefont {Zhang}}, \bibinfo {author} {\bibfnamefont
  {F.}~\bibnamefont {Casola}}, \bibinfo {author} {\bibfnamefont
  {T.}~\bibnamefont {Taniguchi}}, \bibinfo {author} {\bibfnamefont
  {K.}~\bibnamefont {Watanabe}}, \bibinfo {author} {\bibfnamefont {M.~M.}\
  \bibnamefont {Fogler}}, \bibinfo {author} {\bibfnamefont {P.}~\bibnamefont
  {Kim}}, \bibinfo {author} {\bibfnamefont {A.}~\bibnamefont {Yacoby}},\ and\
  \bibinfo {author} {\bibfnamefont {R.~L.}\ \bibnamefont {Walsworth}},\
  }\bibfield  {title} {\bibinfo {title} {Imaging viscous flow of the {Dirac}
  fluid in graphene},\ }\href {https://doi.org/10.1038/s41586-020-2507-2}
  {\bibfield  {journal} {\bibinfo  {journal} {Nature}\ }\textbf {\bibinfo
  {volume} {583}},\ \bibinfo {pages} {537} (\bibinfo {year}
  {2020})}\BibitemShut {NoStop}%
\bibitem [{\citenamefont {Thiel}\ \emph {et~al.}(2019)\citenamefont {Thiel},
  \citenamefont {Wang}, \citenamefont {Tschudin}, \citenamefont {Rohner},
  \citenamefont {Gutiérrez-Lezama}, \citenamefont {Ubrig}, \citenamefont
  {Gibertini}, \citenamefont {Giannini}, \citenamefont {Morpurgo},\ and\
  \citenamefont {Maletinsky}}]{thiel_probing_2019}%
  \BibitemOpen
  \bibfield  {author} {\bibinfo {author} {\bibfnamefont {L.}~\bibnamefont
  {Thiel}}, \bibinfo {author} {\bibfnamefont {Z.}~\bibnamefont {Wang}},
  \bibinfo {author} {\bibfnamefont {M.~A.}\ \bibnamefont {Tschudin}}, \bibinfo
  {author} {\bibfnamefont {D.}~\bibnamefont {Rohner}}, \bibinfo {author}
  {\bibfnamefont {I.}~\bibnamefont {Gutiérrez-Lezama}}, \bibinfo {author}
  {\bibfnamefont {N.}~\bibnamefont {Ubrig}}, \bibinfo {author} {\bibfnamefont
  {M.}~\bibnamefont {Gibertini}}, \bibinfo {author} {\bibfnamefont
  {E.}~\bibnamefont {Giannini}}, \bibinfo {author} {\bibfnamefont {A.~F.}\
  \bibnamefont {Morpurgo}},\ and\ \bibinfo {author} {\bibfnamefont
  {P.}~\bibnamefont {Maletinsky}},\ }\bibfield  {title} {\bibinfo {title}
  {Probing magnetism in {2D} materials at the nanoscale with single-spin
  microscopy},\ }\href {https://doi.org/10.1126/science.aav6926} {\bibfield
  {journal} {\bibinfo  {journal} {Science}\ }\textbf {\bibinfo {volume}
  {364}},\ \bibinfo {pages} {973} (\bibinfo {year} {2019})}\BibitemShut
  {NoStop}%
\bibitem [{\citenamefont {Abobeih}\ \emph {et~al.}(2019)\citenamefont
  {Abobeih}, \citenamefont {Randall}, \citenamefont {Bradley}, \citenamefont
  {Bartling}, \citenamefont {Bakker}, \citenamefont {Degen}, \citenamefont
  {Markham}, \citenamefont {Twitchen},\ and\ \citenamefont
  {Taminiau}}]{abobeih_atomic-scale_2019}%
  \BibitemOpen
  \bibfield  {author} {\bibinfo {author} {\bibfnamefont {M.~H.}\ \bibnamefont
  {Abobeih}}, \bibinfo {author} {\bibfnamefont {J.}~\bibnamefont {Randall}},
  \bibinfo {author} {\bibfnamefont {C.~E.}\ \bibnamefont {Bradley}}, \bibinfo
  {author} {\bibfnamefont {H.~P.}\ \bibnamefont {Bartling}}, \bibinfo {author}
  {\bibfnamefont {M.~A.}\ \bibnamefont {Bakker}}, \bibinfo {author}
  {\bibfnamefont {M.~J.}\ \bibnamefont {Degen}}, \bibinfo {author}
  {\bibfnamefont {M.}~\bibnamefont {Markham}}, \bibinfo {author} {\bibfnamefont
  {D.~J.}\ \bibnamefont {Twitchen}},\ and\ \bibinfo {author} {\bibfnamefont
  {T.~H.}\ \bibnamefont {Taminiau}},\ }\bibfield  {title} {\bibinfo {title}
  {Atomic-scale imaging of a 27-nuclear-spin cluster using a quantum sensor},\
  }\href {https://doi.org/10.1038/s41586-019-1834-7} {\bibfield  {journal}
  {\bibinfo  {journal} {Nature}\ }\textbf {\bibinfo {volume} {576}},\ \bibinfo
  {pages} {411} (\bibinfo {year} {2019})}\BibitemShut {NoStop}%
\bibitem [{\citenamefont {McLaughlin}\ \emph {et~al.}(2021)\citenamefont
  {McLaughlin}, \citenamefont {Wang}, \citenamefont {Huang}, \citenamefont
  {Lee-Wong}, \citenamefont {Hu}, \citenamefont {Lu}, \citenamefont {Yan},
  \citenamefont {Gu}, \citenamefont {Wu}, \citenamefont {You},\ and\
  \citenamefont {Du}}]{mclaughlin_strong_2021}%
  \BibitemOpen
  \bibfield  {author} {\bibinfo {author} {\bibfnamefont {N.~J.}\ \bibnamefont
  {McLaughlin}}, \bibinfo {author} {\bibfnamefont {H.}~\bibnamefont {Wang}},
  \bibinfo {author} {\bibfnamefont {M.}~\bibnamefont {Huang}}, \bibinfo
  {author} {\bibfnamefont {E.}~\bibnamefont {Lee-Wong}}, \bibinfo {author}
  {\bibfnamefont {L.}~\bibnamefont {Hu}}, \bibinfo {author} {\bibfnamefont
  {H.}~\bibnamefont {Lu}}, \bibinfo {author} {\bibfnamefont {G.~Q.}\
  \bibnamefont {Yan}}, \bibinfo {author} {\bibfnamefont {G.}~\bibnamefont
  {Gu}}, \bibinfo {author} {\bibfnamefont {C.}~\bibnamefont {Wu}}, \bibinfo
  {author} {\bibfnamefont {Y.-Z.}\ \bibnamefont {You}},\ and\ \bibinfo {author}
  {\bibfnamefont {C.~R.}\ \bibnamefont {Du}},\ }\bibfield  {title} {\bibinfo
  {title} {Strong {Correlation} {Between} {Superconductivity} and
  {Ferromagnetism} in an {Fe}-{Chalcogenide} {Superconductor}},\ }\href
  {https://doi.org/10.1021/acs.nanolett.1c02424} {\bibfield  {journal}
  {\bibinfo  {journal} {Nano Letters}\ }\textbf {\bibinfo {volume} {21}},\
  \bibinfo {pages} {7277} (\bibinfo {year} {2021})}\BibitemShut {NoStop}%
\bibitem [{\citenamefont {Berry}(2009)}]{berry_optical_2009}%
  \BibitemOpen
  \bibfield  {author} {\bibinfo {author} {\bibfnamefont {M.~V.}\ \bibnamefont
  {Berry}},\ }\bibfield  {title} {\bibinfo {title} {Optical currents},\ }\href
  {https://doi.org/10.1088/1464-4258/11/9/094001} {\bibfield  {journal}
  {\bibinfo  {journal} {Journal of Optics A: Pure and Applied Optics}\ }\textbf
  {\bibinfo {volume} {11}},\ \bibinfo {pages} {094001} (\bibinfo {year}
  {2009})}\BibitemShut {NoStop}%
\bibitem [{\citenamefont {Kalhor}\ \emph {et~al.}(2016)\citenamefont {Kalhor},
  \citenamefont {Thundat},\ and\ \citenamefont
  {Jacob}}]{kalhor_universal_2016}%
  \BibitemOpen
  \bibfield  {author} {\bibinfo {author} {\bibfnamefont {F.}~\bibnamefont
  {Kalhor}}, \bibinfo {author} {\bibfnamefont {T.}~\bibnamefont {Thundat}},\
  and\ \bibinfo {author} {\bibfnamefont {Z.}~\bibnamefont {Jacob}},\ }\bibfield
   {title} {\bibinfo {title} {Universal spin-momentum locked optical forces},\
  }\href {https://doi.org/10.1063/1.4941539} {\bibfield  {journal} {\bibinfo
  {journal} {Applied Physics Letters}\ }\textbf {\bibinfo {volume} {108}},\
  \bibinfo {pages} {061102} (\bibinfo {year} {2016})}\BibitemShut {NoStop}%
\bibitem [{\citenamefont {Yang}\ \emph {et~al.}(2020)\citenamefont {Yang},
  \citenamefont {Khosravi},\ and\ \citenamefont {Jacob}}]{yang_quantum_2020}%
  \BibitemOpen
  \bibfield  {author} {\bibinfo {author} {\bibfnamefont {L.-P.}\ \bibnamefont
  {Yang}}, \bibinfo {author} {\bibfnamefont {F.}~\bibnamefont {Khosravi}},\
  and\ \bibinfo {author} {\bibfnamefont {Z.}~\bibnamefont {Jacob}},\ }\bibfield
   {title} {\bibinfo {title} {Quantum spin operator of the photon},\ }\href
  {http://arxiv.org/abs/2004.03771} {\bibfield  {journal} {\bibinfo  {journal}
  {arXiv:2004.03771}\ } (\bibinfo {year} {2020})}\BibitemShut {NoStop}%
\bibitem [{\citenamefont {Taylor}\ \emph {et~al.}(2008)\citenamefont {Taylor},
  \citenamefont {Cappellaro}, \citenamefont {Childress}, \citenamefont {Jiang},
  \citenamefont {Budker}, \citenamefont {Hemmer}, \citenamefont {Yacoby},
  \citenamefont {Walsworth},\ and\ \citenamefont
  {Lukin}}]{taylor_high-sensitivity_2008}%
  \BibitemOpen
  \bibfield  {author} {\bibinfo {author} {\bibfnamefont {J.~M.}\ \bibnamefont
  {Taylor}}, \bibinfo {author} {\bibfnamefont {P.}~\bibnamefont {Cappellaro}},
  \bibinfo {author} {\bibfnamefont {L.}~\bibnamefont {Childress}}, \bibinfo
  {author} {\bibfnamefont {L.}~\bibnamefont {Jiang}}, \bibinfo {author}
  {\bibfnamefont {D.}~\bibnamefont {Budker}}, \bibinfo {author} {\bibfnamefont
  {P.~R.}\ \bibnamefont {Hemmer}}, \bibinfo {author} {\bibfnamefont
  {A.}~\bibnamefont {Yacoby}}, \bibinfo {author} {\bibfnamefont
  {R.}~\bibnamefont {Walsworth}},\ and\ \bibinfo {author} {\bibfnamefont
  {M.~D.}\ \bibnamefont {Lukin}},\ }\bibfield  {title} {\bibinfo {title}
  {High-sensitivity diamond magnetometer with nanoscale resolution},\ }\href
  {https://doi.org/10.1038/nphys1075} {\bibfield  {journal} {\bibinfo
  {journal} {Nature Physics}\ }\textbf {\bibinfo {volume} {4}},\ \bibinfo
  {pages} {810} (\bibinfo {year} {2008})}\BibitemShut {NoStop}%
\bibitem [{\citenamefont {Naydenov}\ \emph {et~al.}(2011)\citenamefont
  {Naydenov}, \citenamefont {Dolde}, \citenamefont {Hall}, \citenamefont
  {Shin}, \citenamefont {Fedder}, \citenamefont {Hollenberg}, \citenamefont
  {Jelezko},\ and\ \citenamefont {Wrachtrup}}]{naydenov_dynamical_2011}%
  \BibitemOpen
  \bibfield  {author} {\bibinfo {author} {\bibfnamefont {B.}~\bibnamefont
  {Naydenov}}, \bibinfo {author} {\bibfnamefont {F.}~\bibnamefont {Dolde}},
  \bibinfo {author} {\bibfnamefont {L.~T.}\ \bibnamefont {Hall}}, \bibinfo
  {author} {\bibfnamefont {C.}~\bibnamefont {Shin}}, \bibinfo {author}
  {\bibfnamefont {H.}~\bibnamefont {Fedder}}, \bibinfo {author} {\bibfnamefont
  {L.~C.~L.}\ \bibnamefont {Hollenberg}}, \bibinfo {author} {\bibfnamefont
  {F.}~\bibnamefont {Jelezko}},\ and\ \bibinfo {author} {\bibfnamefont
  {J.}~\bibnamefont {Wrachtrup}},\ }\bibfield  {title} {\bibinfo {title}
  {Dynamical decoupling of a single-electron spin at room temperature},\ }\href
  {https://doi.org/10.1103/PhysRevB.83.081201} {\bibfield  {journal} {\bibinfo
  {journal} {Physical Review B}\ }\textbf {\bibinfo {volume} {83}},\ \bibinfo
  {pages} {081201} (\bibinfo {year} {2011})}\BibitemShut {NoStop}%
\bibitem [{\citenamefont {Childress}\ \emph {et~al.}(2006)\citenamefont
  {Childress}, \citenamefont {Gurudev~Dutt}, \citenamefont {Taylor},
  \citenamefont {Zibrov}, \citenamefont {Jelezko}, \citenamefont {Wrachtrup},
  \citenamefont {Hemmer},\ and\ \citenamefont
  {Lukin}}]{childress_coherent_2006}%
  \BibitemOpen
  \bibfield  {author} {\bibinfo {author} {\bibfnamefont {L.}~\bibnamefont
  {Childress}}, \bibinfo {author} {\bibfnamefont {M.~V.}\ \bibnamefont
  {Gurudev~Dutt}}, \bibinfo {author} {\bibfnamefont {J.~M.}\ \bibnamefont
  {Taylor}}, \bibinfo {author} {\bibfnamefont {A.~S.}\ \bibnamefont {Zibrov}},
  \bibinfo {author} {\bibfnamefont {F.}~\bibnamefont {Jelezko}}, \bibinfo
  {author} {\bibfnamefont {J.}~\bibnamefont {Wrachtrup}}, \bibinfo {author}
  {\bibfnamefont {P.~R.}\ \bibnamefont {Hemmer}},\ and\ \bibinfo {author}
  {\bibfnamefont {M.~D.}\ \bibnamefont {Lukin}},\ }\bibfield  {title} {\bibinfo
  {title} {Coherent {Dynamics} of {Coupled} {Electron} and {Nuclear} {Spin}
  {Qubits} in {Diamond}},\ }\href {https://doi.org/10.1126/science.1131871}
  {\bibfield  {journal} {\bibinfo  {journal} {Science}\ }\textbf {\bibinfo
  {volume} {314}},\ \bibinfo {pages} {281} (\bibinfo {year}
  {2006})}\BibitemShut {NoStop}%
\end{thebibliography}%

\onecolumngrid

\renewcommand{\thefigure}{S\arabic{figure}}
\setcounter{figure}{0}

\section*{Supplementary information}
\subsection*{
Antenna fabrication}
The antennas were fabricated on commercially available optical grade diamond plates grown by chemical vapor deposition (CVD) from elementsix\texttrademark{} of dimensions 4.5 mm by 4.5 mm, with orientation $\langle 100 \rangle$. Samples were first soaked in piranha and nitric acid each for 30 minutes, then rinsed in distilled water. A solvent cleaning procedure involving toluene, acetone, and isopropanol each for 5 minutes was done to dissolve any inorganic contaminants on the diamond plates. A stress-release etch of $5\mu m$ on either side of the plate was done in Plasma-Therm Apex SLR inductively coupled plasma (ICP) etcher using Ar, O2, and Cl2 chemistry. Lithography was then done on the samples by spinning CSAR electron beam resist at a speed of 1000 rpm followed by baking at 150°C on a hot plate. The patterns were then exposed in JEOL JBX-8100 FS E beam writer using a current of $30 pA$. After exposure, the plates were developed in xylene for 60 seconds and stopped in isopropanol for another minute. A descum process was then done at 110W using Ar/O2 plasma for 1 minute before metal deposition using electron beam metal deposition system at a base pressure in 1e-6 Torr. 30 nm/300 nm of Ti/Au was deposited. Samples were then lifted off in acetone. 

\begin{figure}[tbh!]
\centering\includegraphics[width=120mm]{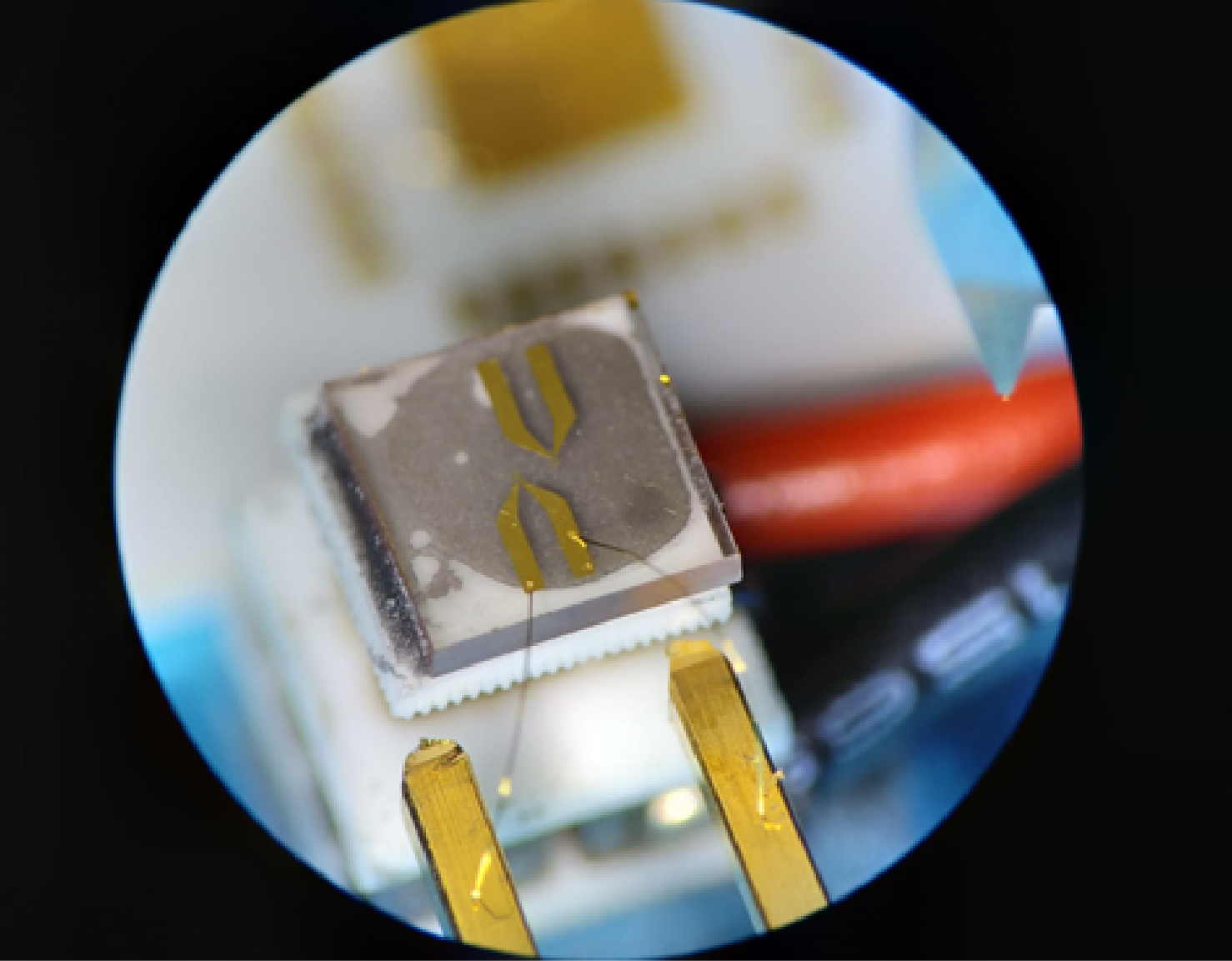}
\caption{Microscope image of the fabricated antenna wire bonded to metallic contacts. The diamond is attached to a thermoelectric cooler.}
\end{figure}

\end{document}